# CONVCONCATNET: A DEEP CONVOLUTIONAL NEURAL NETWORK TO RECONSTRUCT MEL SPECTROGRAM FROM THE EEG

*Xiran Xu [1,3], Bo Wang [1,3], Yujie Yan [2,3], Haolin Zhu [1,3], Zechen Zhang, [1,3], Xihong Wu [1,3], Jing Chen [1,2,3]*

[1] Speech and Hearing Research Center, School of Intelligence Science and Technology, Peking University
[2] National Biomedical Imaging Center, College of Future Technology, Peking University
[3] National Key Laboratory of General Artificial Intelligence, China
janechenjing@pku.edu.cn

## ABSTRACT

To investigate the processing of speech in the brain, simple linear models are commonly used to establish a relationship between brain signals and speech features. However, these linear models are ill-equipped to model a highly dynamic and complex non-linear system like the brain. Although non-linear methods with neural networks have been developed recently, reconstructing unseen stimuli from unseen subjects' EEG is still a highly challenging task. This work presents a novel method, ConvConcatNet, to reconstruct mel-specgrams from EEG, in which the deep convolution neural network and extensive concatenation operation were combined. With our ConvConcatNet model, the Pearson correlation between the reconstructed and the target mel-spectrogram can achieve 0.0420, which was ranked as No.1 in the Task 2 of the Auditory EEG Challenge. The codes and models to implement our work will be available on Github:
https://github.com/xuxiran/ConvConcatNet

## 1. INTRODUCTION

Reconstructing speech from brain activity using non-invasive recordings is a long-awaited goal in both healthcare and neuroscience [1]. However, the current technology is still far from achieving this goal. Though methods for reconstructing speech envelope from EEG have been developed quickly in recent years [2], studies aiming to reconstruct mel-spectrograms are quite rare. On the other side, compared to the most subject-dependent methods, the EEG decoding method which can be used to reconstruct unseen stimuli from the EEG of unseen subjects is closer to practical application, but it is more challenging. Here, "unseen" means the type of data was not included in both the training and validation datasets. For these reasons, it is valuable to discuss the feasibility of reconstructing the mel-spectrogram of unseen stimuli from the EEG of unseen subjects. This task was defined as TASK2 in the Auditory EEG Challenge of ICASSP 2024.

The difficulty of TASK2 includes: 1) there is currently insufficient evidence to suggest that low-frequency neural activity, which can be better recorded by EEG compared to high-frequency neural activities, closely tracks the mel-spectrogram of speech; 2) EEG recordings are quite noisy signals and the noise vary greatly across sessions and individuals [1]. To address these difficulties, three solutions were proposed in this work:

(1) An iterative deep convolutional neural network was designed as the model architecture. Deep convolutional neural networks have been proven effective in extracting spatial-temporal features from EEG [2], [3]. On the other side, extensive concatenation operations were used to iteratively fuse EEG features at different levels.
(2) The envelope was used to help the model learn mel-spectrogram features. Neural responses have been proven to track the speech envelope [2]. Considering the similarity between the envelope and the mel-spectrogram, it is expected that providing the envelope during the training stage can help the model learn the mel-spectrogram features.
(3) A specific data-splitting strategy was used for validation to avoid overfitting. The validation set only consisted of unseen stimuli and evoked EEG from unseen subjects to avoid overfitting the features of specific subjects or stimuli.

## 2. METHODS

### 2.1. Datasets

The challenge's training dataset [4] provides the EEG responses with the corresponding stimuli of 85 young normal-hearing subjects. The testing dataset provides EEG responses without the corresponding stimuli of other 20 young normal-hearing subjects. Only 64 Hz preprocessed EEG data was used in this work. A detailed introduction to the dataset can be found in [4].

### 2.2. Model

An iterative deep convolutional neural network inspired by the previous study [2] called ConvConcatNet was proposed. As shown in Figure 1a, ConvConcatNet consists of 6 blocks, and each block consists of 4 different parts (Figure 1b). The first part of each block is the CNN stack consisting of 5 convolutional layers. The second part is a simple fully connected linear layer. The third part is the output context layer consisting of a zero-padding, a temporal convolutional layer, LeakyReLU activation function, and layer normalization. The last part is a sample spatial attention layer.

As shown in Figure 1c, each of the first four layers in the CNN stack contains Sconv and Tconv. Sconv is a pointwise convolution followed by LLP (Layer normalization, LeakyReLU activation function, and zero-padding). Tconv is a temporal convolution with groups equal to the input channel number followed by LLP. The last layer in the CNN stack, conv, contains a sample temporal convolutional layer followed by LLP.

This work is supported by the National Key Research and Development Program of China (No.2021ZD0201503), National Natural Science Foundation of China (No.12074012), and High-performance Computing Platform of Peking University.

Since the extensive use of concatenation operations, this model was named as ConvConcatNet. As shown in Figures 1a and 1b, the EEG, the output of the context layer, and the output of the spatial attention layer are concatenated along the channel dimension and used as the initial input for the next block. Additionally, in the CNN stack, the output of Sconv is concatenated with the initial input along the channel dimension.

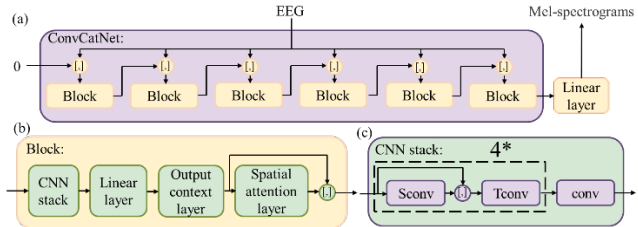

Fig. 1. Structure of the proposed ConvConcatNet. (a) the overall network architecture of ConvConcatNet. (b) four different parts in each block. (c) the network architecture of the CNN stack.

### 2.3. Training and validation

The envelope was used to help the model learn mel-spectrogram features. In the training stage, the envelope and a 10-subband-mel-spectrogram were concatenated along the subband dimension to form 11-subband data. In the validation stage, only the last 10 subbands corresponding to the mel-spectrogram were reserved to calculate the correlation coefficient with the target mel-spectrogram.

The EEG responses in the testing set were recorded from unseen subjects. Besides, all stimuli except AB1 (audiobook 1) [4] were also unseen in training dataset. To avoid overfitting the features of specific subjects or stimuli, it is necessary to ensure each stimulus and its corresponding EEG seen in the training set is unseen in the validation set. A four-fold cross-validation approach was employed, wherein AB1 was excluded from the validation set in each fold. This decision was made because AB1 had already been included in the training set. The detailed data-splitting is shown in Table 1 and all the results reported in the next section are shown across these four folds.

Tab. 1. Detailed data-splitting of cross-validation

| Fold | Subjects in training set | Subjects in validation set |
|------|--------------------------|----------------------------|
| 1 | 27-85 | 1-26 |
| 2 | 1-26,49-85 | 27-48 |
| 3 | 1-48,72-85 | 49-71 |
| 4 | 1-71 | 72-85 |

To improve model performance, ensemble learning was employed. Specifically, a total of 130 models were trained on each fold using different random seeds, resulting in a total of 520 models. The prediction was made for each model and the predictions from all models were normalized and then averaged to get the final prediction. For local evaluation purposes, the predictions from each fold were also normalized and averaged to verify the effectiveness of the ensembling.

The neural networks were implemented with the Pytorch and trained on an A100 GPU. The Adaptive Moment Estimation (Adam) optimizer was employed to maximize the correlation with a learning rate of $10^{-3}$.

### 3. RESULTS AND DISCUSSION

The results in each fold of the four models will be reported. The first model is VLAAI (Very Large Augmented Auditory Inference) [2], which is the baseline nonlinear model of this challenge. To validate the effectiveness of providing envelopes in the training stage, the second model is ConvConcatNet without providing envelopes (i.e., ConvConcatNet w/o env.) in the training stage. The third model is ConvConcatNet, and the last is ConvConcatNet ensembled.

Tab. 2. Results in each fold of four models, measured by Pearson correlation between the reconstructed and the target mel-spectrogram. The results were averaged over 10 subbands.

| Model | Fold 1 | Fold 2 | Fold 3 | Fold 4 |
|-------|--------|--------|--------|--------|
| VLAAI | 0.0505 | 0.0548 | 0.0416 | 0.0550 |
| ConvConcatNet w/o env. | 0.0538 | 0.0599 | 0.0460 | 0.0615 |
| ConvConcatNet | 0.0586 | 0.0604 | 0.0489 | 0.0639 |
| ConvConcatNet ensembled | **0.0627** | **0.0663** | **0.0536** | **0.0703** |

As shown in Table 2, ConvConcatNet outperforms VLAAI, indicating the benefits of concatenating EEG and the extracted features iteratively. Additionally, the manipulation of adding envelope as features during the training stage and model ensembling can improve performance.

Although the performance of ConvConcatNet outperformed the other models in this competition, the limitations need to be addressed: 1) the effectiveness of splitting the training dataset into training and validation sets needs to be verified when the target mel-spectrogram of the testing set is released to the public; 2) considering the high correlation between the envelope and the mel-spectrogram, further experimental analysis is required to determine whether the current model indeed reconstructs specific features of the mel-spectrogram or only the shared features between the envelope and the mel-spectrogram; 3) the current correlation coefficient for the reconstructed mel-spectrogram is still low, the bigger database and the better models are worthy to be developed in future.

### 4. CONCLUSION

In this work, a novel ConvConcatNet method was proposed, and the evaluation experiments confirmed the effectiveness of the model structure and the training strategy. Although the performance of our model was ranked as No.1 in the competition, the accuracy of reconstructing mel-spectrograms from EEG is still low, and the task remains challenging.